# A Fourier-based Solving Approach for the Transport of Intensity Equation without Typical Restrictions

**SOHEIL MEHRABKHANI,**[1,2,*] **LENNART WEFELNBERG,**[1] **AND THOMAS SCHNEIDER**[1,2]

[1]*Terahertz-Photonics Group, Institut für Hochfrequenztechnik, TU Braunschweig, Schleinitzstraße 22, 38106 Braunschweig, Germany*
[2]*Lena, Laboratory for Emerging Nanometrology, Braunschweig, Germany*
*\*soheil.mehrabkhani@ihf.tu-bs.de*

**Abstract:** The Transport-of-Intensity equation (TIE) has been proven as a standard approach for phase retrieval. Some high efficiency solving methods for the TIE, extensively used in many works, are based on a Fourier-Transform (FT). However, to solve the TIE by these methods several assumptions have to be made. A common assumption is that there are no zero values for the intensity distribution allowed. The two most widespread Fourier-based approaches have further restrictions. One of these requires the uniformity of the intensity distribution and the other assumes the collinearity of the intensity and phase gradients. In this paper, we present an approach, which does not need any of these assumptions and consequently extends the application domain of the TIE.

## 1. Introduction

Within the last decades, the number of applications for coherent diffraction has grown enormously. Among them are biomedical imaging, coherent tomography, digital holography, coherent x-ray imaging, holographic microscopy, optical metrology, wave sensing and many more. The information of the coherent illumination of an object is contained in a complex scalar field, composed of intensity and phase. Although, the intensity distribution can be measured by a CCD camera, the phase measurement is a rather challenging task. As a standard technique, interferometry has been used. However, it requires a cumbersome very sensible experimental setup. Due to powerful computer processors, available in recent years, phase retrieval has been established as the most widely used alternative to interferometry. The main advantage of phase retrieval is its ability to extract the phase information from the intensity measurements by appropriate numerical algorithms. Until now, several phase retrieval algorithms have been developed. However, each method has its drawbacks and limitations, which restrict its application. One of the algorithms in the paraxial domain is based on the Transport-of-Intensity equation (TIE) originally developed by Teague [1]. Contrary to iterative approaches like the Gerchberg-Saxton algorithm and its numerous modifications, the TIE is a deterministic approach. Thus, it avoids problems like non-convergence or stagnation, which are typical in many iterative approaches. Although, it is an elegant deterministic way for phase retrieval, its solution is not trivial. Two very successful fast solving approaches are based on the Fast Fourier Transform (FFT). However, they suffer from different assumptions related to the intensity behavior, which automatically restrict the application of these tools. One of these approaches proposed by [2] and applied in [3-20], for instance, assumes the uniformity of the intensity distribution (orthogonal to the propagation direction). Another approach [21-32], is also restricted to very special cases due to the assumption that the transverse gradient of the phase and the intensity must be collinear [33]. Moreover, zero values for the intensity must be strictly avoided in both approaches. In this work, we present an efficient algorithm without these restrictions, making the applicability of the TIE more general.

## 2.  Iterative solution for the TIE

The TIE describes the coupling between the intensity $I$ and the phase $\phi$ in a plane orthogonal to the direction of the paraxial wave [1]

$$\nabla_\perp \cdot (I \nabla_\perp \phi) = -k \frac{\partial I}{\partial z}, \tag{1}$$

where $k$ is the wave number $(k = 2\pi/\lambda)$ and $z$ is the propagation direction of the field. If the complex amplitude is replaced by, $\sqrt{I} \exp(i\phi)$ the TIE is the imaginary part of the paraxial Helmholtz equation [34]. In principle, the two coefficients $k$ and $I$ are known. The axial intensity derivative $\partial I/\partial z$ can be linearly approximated by the Finite Difference method using two intensity measurements in two parallel planes orthogonal to the propagation with the distance $\Delta z$:

$$\frac{\partial I}{\partial z} \simeq \frac{I_2 - I_1}{\Delta z}. \tag{2}$$

In order to remove the aforementioned restrictions of the standard solutions of the TIE, we propose the following change of Eq (1):

$$(I+C)\nabla_\perp^2 \phi = -k \frac{\partial I}{\partial z} - \nabla_\perp I \cdot \nabla_\perp \phi + C \nabla_\perp^2 \phi, \tag{3}$$

where the parameter $C$ is an arbitrary positive constant, $C(x,y) = const$. Since it avoids the division by zero, the constant $C$ enables zero values in the intensity distribution. It should be noted, that the Eqs. (1) and (3) are basically equivalent. Equation (3) is simply derived by adding $C\nabla_\perp^2 \phi$ on both sides of Eq. (1) and expanding the divergence operator. In this general form, Eq. (3) cannot be solved easily, however, an iterative approach may be applied to convert it to a Poisson Equation. Such an iterative Poisson equation additionally avoids the restrictions of the other TIE solving approaches, mentioned above. Let us consider the phase on the right side of Eq. (3) as the phase after the $n$-th iteration, whereas the left side is the next one. This way, the phase on the left and right side can be treated as two different mathematical variables $\phi_n$ and $\phi_{n+1}$ in an iterative approach. Consequently, Eq. (3) can be written as a Poisson equation:

$$\nabla_{\perp}^{2}\phi_{n+1} = \frac{-k\frac{\partial I}{\partial z} - \nabla_{\perp}I \cdot \nabla_{\perp}\phi_{n} + C\nabla_{\perp}^{2}\phi_{n}}{I+C}. \quad (4)$$

To solve equation (4), the Fourier transform (FT) can be used:

$$-4\pi^{2}(f_{x}^{2}+f_{y}^{2})\tilde{\phi}_{n+1} = FT\left\{\frac{-k\frac{\partial I}{\partial z} - \nabla_{\perp}I \cdot \nabla_{\perp}\phi_{n} + C\nabla_{\perp}^{2}\phi_{n}}{I+C}\right\}. \quad (5)$$

For all frequency pairs $(f_x, f_y) \neq 0$ Eq. (5) can be rewritten as:

$$\tilde{\phi}_{n+1} = \frac{1}{-4\pi^{2}(f_{x}^{2}+f_{y}^{2})} FT\left\{\frac{-k\frac{\partial I}{\partial z} - \nabla_{\perp}I \cdot \nabla_{\perp}\phi_{n} + C\nabla_{\perp}^{2}\phi_{n}}{I+C}\right\}. \quad (6)$$

The singularity at the frequency point $(f_x, f_y) = 0$ may be simply removed by a constant predefined value for the Fourier- Transform of the phase at the frequency center $\tilde{\phi}_{n+1}(0,0) = 0$. This corresponds to a constant phase shift in the spatial domain, which has no influence on the result. Eq. (6) provides $\tilde{\phi}_{n+1}(f_x, f_y)$, which is the FT of $\phi_{n+1}(x, y)$. Finally, the phase $\phi_{n+1}(x, y)$ is calculated by the inverse Fourier Transform (IFT):

$$\phi_{n+1} = \frac{-1}{4\pi^{2}} IFT\left\{\frac{1}{f_{x}^{2}+f_{y}^{2}} FT\left\{\frac{-k\frac{\partial I}{\partial z} - \nabla_{\perp}I \cdot \nabla_{\perp}\phi_{n} + C\nabla_{\perp}^{2}\phi_{n}}{I+C}\right\}\right\}. \quad (7)$$

The initial phase is set to $\phi_0(x, y) = 0$. In order to get the next estimation for the phase $\phi_{n+1}$, in each iteration the phase $\phi_n$ is introduced into Eq. (7). The iteration will be repeated, until the convergence is reached. The termination condition for the algorithm may be determined by a fixed iteration number, for instance.

### 3. Numerical Implementation

*3.1 Constraints on the phase changes under the paraxial assumption*

The TIE is only valid in the paraxial domain and consequently its solution must satisfy the paraxial condition. However, in the proposed iterative method, the start phase distribution $\phi_0(x, y) = 0$ and subsequently calculated phase distributions $\phi_n(x, y)$ are not necessarily a solution of the TIE. They are just approximations for the correct solution, which must be improved in each iteration. Thus, these estimations may violate the paraxial assumption. Consequently, some a priori information concerning the paraxial domain must be included as additional constraints to prevent a non-physical solution. In principle, there is already a well-known condition related to $\nabla_\perp \phi$, which is a result of the Eikonal equation [34]:

$$\nabla_\perp \phi \simeq \vec{k}_\perp, \tag{8}$$

where $\vec{k}_\perp$ is the transverse wave vector. Due to the paraxial assumption, the vertical component of the wave number is proportional to the vertical position [35]. Thus, the inequality $k_\perp \leq NA\, k$ is satisfied. The numerical aperture $NA$ of the images recorded by the CCD camera can be calculated from simple geometrical considerations. Within the paraxial assumption, Eq. (8) results in a paraxial constraint for the phase gradient $|\nabla_\perp \phi| \leq NA\, k$.

The right side of Eq. (7) depends on both $\nabla_\perp \phi$ and $\nabla_\perp^2 \phi$ and thus, the paraxial condition concerning $\nabla_\perp^2 \phi$ must be derived as well. According to the definition of the Laplace operator including its discretization with the spacing $\Delta x$, it follows:

$$\left|\nabla_\perp^2 \phi\right| = \left|\frac{\partial^2 \phi}{\partial x^2} + \frac{\partial^2 \phi}{\partial y^2}\right| \simeq \frac{1}{\Delta x}\left|\frac{\partial \phi(x+\Delta x)}{\partial x} - \frac{\partial \phi(x)}{\partial x} + \frac{\partial \phi(y+\Delta x)}{\partial y} - \frac{\partial \phi(y)}{\partial y}\right|. \tag{9}$$

From Eq. (8), the following relationships can be derived:

$$k_{1,x} \simeq \frac{\partial \phi(x)}{\partial x}, \quad k_{1,y} \simeq \frac{\partial \phi(y)}{\partial y}, \quad k_{2,x} \simeq \frac{\partial \phi(x+\Delta x)}{\partial x}, \quad k_{2,y} \simeq \frac{\partial \phi(y+\Delta y)}{\partial y}, \tag{10}$$

which will be used to replace the phase derivatives in Eq. (9) by the wave vector components. The triangle inequality [36] within the Eqs. (9) and (10) provide a very useful estimation for the maximum of $|\nabla_\perp^2 \phi|$:

$$\left|\nabla_\perp^2 \phi\right| \simeq \frac{1}{\Delta x}\left|k_{2,x} - k_{1,x} + k_{2,y} - k_{1,y}\right| \leq \frac{1}{\Delta x}\left(\left|k_{1,x} + k_{1,y}\right| + \left|k_{2,x} + k_{2,y}\right|\right). \tag{11}$$

Eq. (10) and the paraxial condition determine a maximum for $|k_x + k_y|$:

$$\left|\nabla_\perp \phi\right|^2 \simeq k_x^2 + k_y^2 \leq NA^2 k^2 \Rightarrow |k_x + k_y| \leq \sqrt{NA^2 k^2 + 2 k_x k_y}, \tag{12}$$

which will be applied in the inequality (11):

$$\left|\nabla_\perp^2 \phi\right| \leq \frac{1}{\Delta x}\left(\sqrt{NA^2 k^2 + 2 k_{1,x} k_{1,y}} + \sqrt{NA^2 k^2 + 2 k_{2,x} k_{2,y}}\right). \tag{13}$$

The advantage of inequality (13) is its dependence on the terms $k_{1,x} k_{1,y}$ and $k_{2,x} k_{2,y}$, whose maximum can be simply found. After, defining the function $g(k_x, k_y) = k_x k_y$, it follows for the paraxial condition: $k_x \leq NA\, k, k_y \leq NA\, k$. Consequently, the points $(k_x, k_y)$ are enclosed in a circle with the radius $NA\, k$. Now consider an arbitrary point inside the circle. Obviously, the value of $g(k_x, k_y)$ grows if the point moves radially outwards. Thus, the maximum of $g(k_x, k_y)$ is at the circle boundary. Therefore, the relationship $k_x^2 + k_y^2 = NA^2 k^2$ describes

$g(k_x, k_y)$ as a single variable function $g(k_x) = k_x \sqrt{NA^2 k^2 - k_x^2}$. Now the extremum condition $\frac{dg}{dk_x} = 0$ results in $g_{max} = NA^2 k^2 / 2$, which yields the following condition for the discretized $\left|\nabla_\perp^2 \phi\right|$:

$$\left|\nabla_\perp^2 \phi\right| \leq \frac{2\sqrt{2} NA\, k}{\Delta x}. \tag{14}$$

Eq. (14) is our proposed paraxial constraint on the discretized Laplace operator of the phase. The inequality (14) is only an estimation for the maximum value of the discretized $\left|\nabla_\perp^2 \phi\right|$ under the assumption of the paraxial wave propagation and in the validity domain of the Eikonal equation [37].

### 3.2 Implementation of the paraxial constraints

To fulfill the paraxial condition, both constraints for the gradient and Laplace operator of the phase must be considered in the algorithm. These are the necessary constraints to make the set of possible solutions smaller and to prevent non-physical solutions for the phase distribution $\phi$. This can be accomplished by establishing the upper limit for $\left|\nabla_\perp \phi\right|$ and $\left|\nabla_\perp^2 \phi\right|$ as well as fixing their value in the algorithm, if their maximum values are reached already at some points. However, as long as they satisfy both constraints, the values for other points can be changed freely in each iteration.

One further restriction, which has a significant influence on the phase reconstruction, is related to the value of $C$. Although, from an analytical point of view the value of $C$ can be any arbitrary positive value, the choosing of the numerical value is not trivial. If $C$ is too high, the term $C\nabla_\perp^2 \phi_n$ is too dominant against the term $k \frac{\partial I}{\partial z}$, which suppresses the crucial information in the intensity derivative. On the other hand, too small values for $C$ cause too high values for $\nabla_\perp^2 \phi_1$, especially at points with $I_1 = 0$, which may violate again the paraxial condition. Consequently, a reasonable estimation of the constant $C$ is required. Due to the start phase, $\phi_0(x, y) = 0$, it follows from Eq. (3):

$$(I_1 + C)\nabla_\perp^2 \phi_1 = -k \frac{\partial I_1}{\partial z}. \tag{15}$$

The condition (14) together with Eq. (15) results in:

$$2\sqrt{2} NA |I_1 + C| \geq \Delta x \left|\frac{\partial I_1}{\partial z}\right|. \tag{16}$$

The parameter $C$ must satisfy the inequality (16) for arbitrary intensity values including zero:

$$C \geq \frac{\Delta x \left|\frac{\partial I_1}{\partial z}\right|_{max}}{2\sqrt{2}NA}. \tag{17}$$

This minimum value will be used in the algorithm.

## 4. Results

The approach described in the last sections was applied to the intensity images shown in Fig. 1 (a) and (b). These two intensity distributions have a distance of $\Delta z = 1\mu m$. To show the strength of the method, we have defined another picture (Fig.1 (c)) as the arbitrary phase distribution of Fig.1 (a). In order to show the reliability of the method and the fact that it is no longer restricted by the assumption that the intensity has to be above zero, we have set 1311 pixels in $I_1$ to zero (black regions marked with arrows in Fig.1 (a)). The intensity $I_2$ (Fig.1 (b)) was calculated from $I_1$ (Fig.1 (a)) and the phase distribution (Fig.1(c)) by the Angular Spectrum method [38]. The phase reconstructed from $I_1$ and $I_2$ by the proposed algorithm can be seen in Fig. 1(d).

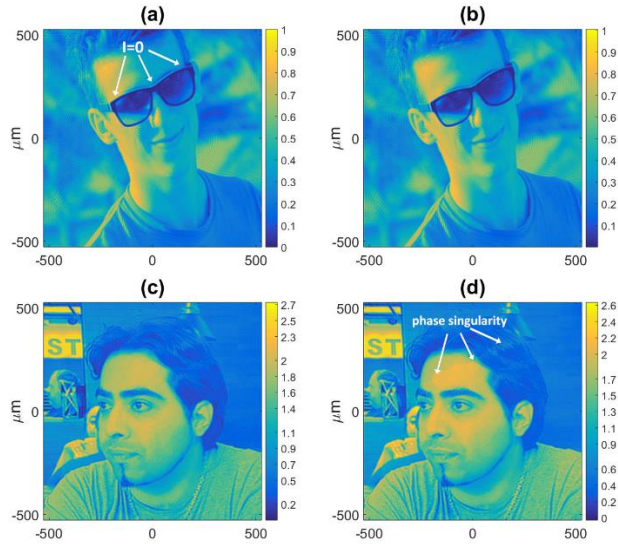

Fig. 1: (a) and (b) are the intensity distributions in the first and the second plane respectively. (c) is the original phase in the first plane and (d) the phase distribution in the first plane reconstructed by the presented method. Both images are quadratic with a width and height of 1.055 mm and the pixel number is 1135 x 1135. The wavelength of the light in the simulation is 633 nm. According to [35], the paraxial condition requires a very small transversal spatial frequency $f_\perp$ comparing to the spatial frequency of the light $f_\perp \ll f$ $(f = 1/\lambda)$. Thus, the maximum transverse spatial frequency of the complex amplitude $u_1 = \sqrt{I_1} \exp(i\phi_1)$ was $f_{\perp max} = 0.05/\lambda$. The used $C$ according Eq. (17) was 0.0382.

In Fig. 1(d) some singularity points can be seen (marked by white arrows). The reason for these phase singularities are the zero values of the corresponding intensity in Fig.1 (a). However, since the phase for an intensity of zero does not make any physical sense, these values are irrelevant. Due to the zero intensity, the corresponding complex amplitude will be zero as well. Thus, the phase value at the singularity points cannot change the complex amplitude (it will always be zero). In general, the singularity points may cause undesired discontinuities in the phase distribution. However, as can be seen in Fig. 1(d), a further advantage of the method is, that the reconstructed phase is almost continuous, even at the singularity points. The convergence of the algorithm is achieved after only 10 iterations, the correlation value of the reconstructed phase with the original phase (Figs. 1(c) and (d)), is 99.6 %, which verifies the reliability and very high efficiency of the proposed algorithm.

## 5. Conclusion

An iterative method for removing the usual intensity restrictions in solving the TIE has been proposed. Contrary to the conventional Fourier-based methods, the intensity must not be homogenous and the intensity and phase gradients must not be collinear. Moreover, the method presented here allows zero values for the intensity. To evaluate the capability of our method, an inhomogenous intensity with a non-collinear gradient of intensity and phase and 1311 points with zero intensity have been used. The validity and capability of the method, has been verified by the achieved correlation coefficient of 99.6 %. The total necessary iteration number was only 10, which results in a very low numerical effort and consequently very high time efficiency.

**Acknowledgments**
We like to thank Janosch Meier and Ali Dorostkar for very fruitful discussions and Stefan Preußler for the proofreading of the manuscript and his support and invaluable suggestions during the writing of the paper.

**Funding**
Niedersächsisches Vorab (NL – 4 Project "QUANOMET") and German Research Foundation (DFG SCHN 716/13-1)